\documentclass[aps,prc,twocolumn,superscriptaddress,showpacs,floatfix,nofootinbib]{revtex4}

\usepackage{amsmath,graphicx}
\usepackage[colorlinks=true,linktocpage=true,linkcolor=blue,citecolor=blue]{hyperref}
\usepackage[usenames,dvipsnames]{color}
\usepackage{float}

\begin{document}

\preprint{}

\title{Relativistic quantum transport coefficients for second-order viscous hydrodynamics}

\author{Wojciech Florkowski}
\affiliation{Institute of Physics, Jan Kochanowski University, PL-25406 Kielce, Poland}
\author{Amaresh Jaiswal}
\affiliation{GSI, Helmholtzzentrum f\"ur Schwerionenforschung, Planckstrasse 1, D-64291 Darmstadt, Germany}
\author{Ewa Maksymiuk}
\affiliation{Institute of Physics, Jan Kochanowski University, PL-25406 Kielce, Poland}
\author{Radoslaw Ryblewski}
\affiliation{The H. Niewodnicza\'nski Institute of Nuclear Physics, Polish Academy of Sciences, PL-31342 Krak\'ow, Poland}
\author{Michael Strickland}
\affiliation{Department of Physics, Kent State University, Kent, Ohio 44242, USA}

\date{\today}

\begin{abstract}

We express the transport coefficients appearing in the second-order 
evolution equations for bulk viscous pressure and shear stress 
tensor using Bose-Einstein, Boltzmann, and Fermi-Dirac statistics 
for the equilibrium distribution function and Grad's 14-moment 
approximation as well as the method of Chapman-Enskog expansion for 
the non-equilibrium part. Specializing to the case of transversally 
homogeneous and boost-invariant longitudinal expansion of the 
viscous medium, we compare the results obtained using the above 
methods with those obtained from the exact solution of the massive 
0+1d relativistic Boltzmann equation in the relaxation-time 
approximation. We show that compared to the 14-moment approximation, 
the hydrodynamic transport coefficients obtained by employing the 
Chapman-Enskog method leads to better agreement with the exact 
solution of the relativistic Boltzmann equation.

\end{abstract}



\maketitle

\section{Introduction}

Relativistic viscous hydrodynamics has been applied quite 
successfully to study and understand various collective phenomena 
observed in the evolution of the strongly interacting QCD matter, 
with very high temperature and density, created in relativistic 
heavy-ion collisions; see Ref.~\cite{Heinz:2013th} for a recent 
review. The derivation of hydrodynamic equations is essentially a 
coarse graining procedure whereby one obtains an effective theory 
describing the long-wavelength low-frequency limit of the 
microscopic dynamics of a system \cite{Heinz:2013th, Landau}. 
Relativistic hydrodynamics is formulated as an order-by-order 
expansion in powers of space-time gradients where ideal 
hydrodynamics is of zeroth order \cite{Heinz:2013th}. The viscous 
effects arising in the first-order theory, also known as the 
relativistic Navier-Stokes theory \cite{Eckart:1940zz, Landau}, 
results in acausal signal propagation and numerical instability. 
While causality is restored in the second-order Israel-Stewart (IS) 
theory \cite{Israel:1979wp}, stability may not be guaranteed \cite 
{Romatschke:2009im}. Consistent formulation of a causal theory of 
relativistic viscous hydrodynamics and accurate determination of the 
associated transport coefficients is currently an active research 
topic \cite {Muronga:2003ta, El:2009vj, Denicol:2012cn, 
Jaiswal:2014isa, Chattopadhyay:2014lya, Denicol:2010xn, 
Jaiswal:2012qm, Jaiswal:2013fc, Bhalerao:2013aha, Denicol:2014vaa, 
Denicol:2014mca, Jaiswal:2013npa, Jaiswal:2013vta, Bhalerao:2013pza, 
Romatschke:2003ms, Florkowski:2010cf, Martinez:2010sc, 
Florkowski:2013lza, Bazow:2013ifa, Nopoush:2014pfa, 
Florkowski:2014bba, Florkowski:2014sfa, Florkowski:2014sda, 
Prakash:1993bt, Wiranata:2012br, Wiranata:2012vv, Davesne:1995ms, 
Wiranata:2014jda}.

The second-order IS equations can be derived in several ways \cite 
{Romatschke:2009im}. For example, in the derivations based on the 
second law of thermodynamics, the hydrodynamic transport 
coefficients related to the relaxation times for bulk and shear 
viscous evolution remain undetermined. While these transport 
coefficients can be obtained in the derivations based on kinetic 
theory \cite{Israel:1979wp, Muronga:2003ta}, the form of 
non-equilibrium phase-space distribution function, $f(x,p)$, has to 
be specified. Two most extensively used methods to determine $f(x,p)$
for a system which is close to local thermodynamic equilibrium are 
(1) the Grad's 14-moment approximation \cite{Grad} and (2) the 
Chapman-Enskog method \cite{Chapman}. Note that while Grad's 
14-moment approximation has been widely employed in the formulation 
of a causal theory of relativistic dissipative hydrodynamics \cite 
{Israel:1979wp, Muronga:2003ta, El:2009vj, Denicol:2010xn, 
Denicol:2012cn, Jaiswal:2012qm, Jaiswal:2013fc, Bhalerao:2013aha, 
Denicol:2014vaa, Denicol:2014mca, Romatschke:2009im}, the 
Chapman-Enskog method remains less explored \cite{Jaiswal:2013npa, 
Jaiswal:2013vta, Bhalerao:2013pza, Jaiswal:2014isa, 
Chattopadhyay:2014lya}. On the other hand, the Chapman-Enskog 
formalism has been often used to extract various transport 
coefficients of hot hadronic matter \cite{Prakash:1993bt, 
Wiranata:2012br, Wiranata:2012vv, Davesne:1995ms, Wiranata:2014jda} 
Although in both methods the distribution function is expanded 
around its equilibrium value $f_0(x,p)$, it has been demonstrated 
that the Chapman-Enskog method in the relaxation-time approximation 
(RTA) leads to better agreement with both microscopic Boltzmann 
simulations as well as exact solutions of the relativistic RTA 
Boltzmann equation \cite {Jaiswal:2013npa, Jaiswal:2013vta, 
Jaiswal:2014isa, Chattopadhyay:2014lya}. This may be attributed to 
the fact that a fixed-order moment expansion, as required in Grad's 
approximation, is not necessary in the Chapman-Enskog method.

Much of the research on the application of viscous hydrodynamics in 
relativistic heavy-ion collisions is devoted to the extraction of 
the shear viscosity to entropy density ratio, $\eta/s$, from the 
analysis of the anisotropic flow data \cite{Romatschke:2007mq, 
Song:2010mg, Schenke:2011bn}. Indeed the estimated $\eta/s$ has been 
found to be close to the conjectured lower bound $\eta/s|_{\rm KSS} 
= 1/4\pi$ \cite{Policastro:2001yc, Kovtun:2004de}. On the other 
hand, a self-consistent and systematic study of the effect of the 
bulk viscosity in numerical simulations of high-energy heavy-ion 
collisions is relatively lacking. This may be attributed to the fact 
that the hot QCD matter is assumed to be nearly conformal and 
therefore the bulk viscosity is estimated to be much smaller 
compared to the shear viscosity. However, in reality, QCD is a 
non-conformal field theory and therefore bulk viscous corrections to 
the energy-momentum tensor should not be ignored in order to 
correctly understand the dynamics of the QCD system. Moreover, for 
the range of temperature explored in heavy-ion collision experiments 
at Relativistic Heavy-Ion Collider (RHIC) and Large Hadron Collider 
(LHC), the magnitude and temperature dependence of the bulk 
viscosity could be large enough to influence the space-time 
evolution of the hot QCD matter \cite {Moore:2008ws, 
Noronha-Hostler:2014dqa, Rose:2014fba, Ryu:2015vwa, Wiranata:2009cz, 
Gavin:1985ph, Chakraborty:2010fr}.

It is important to note that the second-order transport 
coefficients, appearing in the evolution equation for the bulk 
viscous pressure, are less understood compared to those of the shear 
stress tensor. While the relaxation time for the bulk viscous 
evolution can be obtained by using the second law of thermodynamics 
in a kinetic theory set up \cite{Jaiswal:2013fc, Bhalerao:2013aha}, 
this method fails to account for the important coupling of the bulk 
viscous pressure with the shear stress tensor \cite 
{Denicol:2014vaa, Denicol:2014mca, Jaiswal:2014isa}. On the other 
hand, for finite masses and classical Boltzmann distribution, the 
second-order transport coefficients corresponding to bulk viscous 
pressure and shear stress tensor have been obtained by employing the 
Grad's 14-moment approximation \cite{Denicol:2014vaa, 
Denicol:2014mca} as well as the Chapman-Enskog method \cite
{Jaiswal:2014isa}, within a purely kinetic theory framework. 
However, these transport coefficients still remain to be determined 
for quantum statistics, i.e., for Bose-Einstein and Fermi-Dirac 
distribution.

In this paper, we express the transport coefficients appearing in 
the second-order viscous evolution equations with non-vanishing 
masses for Bose-Einstein, Boltzmann and Fermi-Dirac distribution. We 
obtain these transport coefficients using the Grad's 14-moment 
approximation as well as the method of Chapman-Enskog expansion. In 
addition, in the case of one-dimensional scaling expansion of the 
viscous medium, we compare the results obtained using the above 
methods with those obtained from the exact solution of massive 0+1d 
relativistic Boltzmann equation in the relaxation-time approximation 
\cite{Florkowski:2014sda}. We demonstrate that the results obtained 
using the Chapman-Enskog method are in better agreement with the 
exact solution of the RTA Boltzmann equation than those obtained 
using the Grad's 14-moment approximation.


\section{Relativistic hydrodynamics}

In the absence of any conserved charges, i.e., for vanishing 
chemical potential, the hydrodynamic evolution of a system is 
governed by the local conservation of energy and momentum: 
$\partial_\mu T^{\mu\nu}=0$. The energy-momentum tensor, $T^{\mu\nu}$, 
which characterizes the macroscopic state of a system, can be 
expressed in terms of the single-particle phase-space distribution 
function, $f(x,p)$, and tensor decomposed into hydrodynamic degrees 
of freedom \cite{deGroot},
\begin{equation}\label{NTD}
T^{\mu\nu} = \!\int\! dP \, p^\mu p^\nu f(x,p) = \epsilon u^\mu u^\nu 
- (P+\Pi)\Delta^{\mu\nu} + \pi^{\mu\nu}.
\end{equation}
Here $dP\equiv gd^3p/[(2\pi)^3p^0]$ is the invariant momentum-space 
integration measure, $g$ being the degeneracy factor, and $p^\mu$ is 
the particle four-momentum. In the tensor decomposition, $\epsilon$, 
$P$, $\Pi$, and $\pi^{\mu\nu}$ are the energy density, the 
thermodynamic pressure, the bulk viscous pressure, and the shear 
stress tensor, respectively. The projection operator 
$\Delta^{\mu\nu}\equiv g^{\mu\nu}-u^\mu u^\nu$ is constructed such 
that it is orthogonal to the hydrodynamic four-velocity $u^\mu$. The 
metric tensor is assumed to be Minkowskian, 
$g^{\mu\nu}\equiv\mathrm{diag} ({+}1,{-}1,{-}1,{-}1)$, and $u^\mu$ 
is defined in the Landau frame: $T^{\mu\nu} u_\nu=\epsilon u^\mu$.

The energy-momentum conservation equation, $\partial_\mu 
T^{\mu\nu}=0$, when projected along and orthogonal to $u^\mu$ gives 
the evolution equations for $\epsilon$ and $u^\mu$,
\begin{align}\label{evol}
\dot\epsilon + (\epsilon+P+\Pi)\theta - \pi^{\mu\nu}\sigma_{\mu\nu} &= 0,  \nonumber\\
(\epsilon+P+\Pi)\dot u^\alpha - \nabla^\alpha (P+\Pi) + \Delta^\alpha_\nu \partial_\mu \pi^{\mu\nu}  &= 0.
\end{align}
Here we have used the usual notation $\theta\equiv\partial_\mu u^\mu$
for the expansion scalar, $\dot A\equiv u^\mu\partial_\mu A$ for 
the co-moving derivative, $\nabla^\alpha\equiv\Delta^{\mu\alpha} 
\partial_\mu$ for the space-like derivative, and 
$\sigma^{\mu\nu}\equiv(\nabla^\mu u^\nu + \nabla^\nu u^\mu)/2 
-(\theta/3)\Delta^{\mu\nu}$ for the velocity stress tensor. The 
energy density and the thermodynamic pressure can be written in 
terms of the equilibrium phase-space distribution function $f_0$ as
\begin{align}
\epsilon_0 &= u_\mu u_\nu \!\int\! dP \, p^\mu p^\nu f_0, \label{ED}\\
P_0 &= -\frac{1}{3}\Delta_{\mu\nu} \!\int\! dP \, p^\mu p^\nu f_0, \label{TP}
\end{align}
where the equilibrium distribution function for  vanishing chemical 
potential is given by 
\begin{equation}\label{EDF}
f_0 = \frac{1}{\exp(\beta u\cdot p) + a}.
\end{equation}
Here $u\cdot p\equiv u_\mu p^\mu$ and $a=-1,0,1$ for Bose-Einstein, 
Boltzmann, and Fermi-Dirac gas, respectively. The inverse 
temperature, $\beta\equiv1/T$, is determined by the matching 
condition $\epsilon=\epsilon_0$.

For a system close to local thermodynamic equilibrium, the 
non-equilibrium phase-space distribution function can be written as 
$f=f_0+\delta f$, where $\delta f\ll f$. Using Eq.~(\ref {NTD}), the 
bulk viscous pressure, $\Pi$, and the shear stress tensor, 
$\pi^{\mu\nu}$, can be expressed in terms of $\delta f$ as \cite
{deGroot}
\begin{align}
\Pi &= -\frac{1}{3}\Delta_{\alpha\beta} \!\int\! dP \, p^\alpha p^\beta\, \delta f, \label{BVP}\\
\pi^{\mu\nu} &= \Delta^{\mu\nu}_{\alpha\beta} \!\int\! dP \, p^\alpha p^\beta\, \delta f, \label{SST}
\end{align}
where $\Delta^{\mu\nu}_{\alpha\beta}\equiv 
(\Delta^{\mu}_{\alpha}\Delta^{\nu}_{\beta} + 
\Delta^{\mu}_{\beta}\Delta^{\nu}_{\alpha})/2 - 
(1/3)\Delta^{\mu\nu}\Delta_{\alpha\beta}$ is a traceless symmetric 
projection operator orthogonal to $u^\mu$ and $\Delta^{\mu\nu}$. In 
the following, we employ the expressions for $\delta f$ obtained 
using the Grad's 14-moment approximation and the Chapman-Enskog like 
iterative solution of the relativistic Boltzmann equation to obtain 
expressions for the quantum transport coefficients associated with 
viscous evolution.


\section{Viscous corrections to the distribution function}

Precise determination of the form of the non-equilibrium single 
particle phase-space distribution function is one of the central 
problems in statistical physics. For a system close to local 
thermodynamic equilibrium, the problem reduces to determining the 
form of the correction to the equilibrium distribution function. 
Within the framework of relativistic hydrodynamics, the viscous 
corrections to the equilibrium distribution function can be obtained 
from two different methods: (1) the moment method and (2) the 
Chapman-Enskog method. The moment method, more popularly known as 
the Grad's 14-moment ansatz, is based on a Taylor-like expansion of 
the non-equilibrium distribution in powers of momenta. On the other 
hand, the Chapman-Enskog method relies on the solution of the 
Boltzmann equation.

Ignoring dissipation due to particle diffusion, the Grad's 14-moment 
approximation leads to
\begin{align}\label{G14M}
\delta f_{\rm G} =\, & \Big[\left\{E_0 + B_0m^2 + D_0(u\cdot p) - 4B_0(u\cdot p)^2\right\}\Pi \nonumber\\
&+ B_2p^\alpha p^\beta \pi_{\alpha\beta}\Big]f_0 \tilde f_0,
\end{align}
where $\tilde f_0 = 1-af_0$. The coefficients $E_0$, $B_0$, $D_0$ 
and $B_2$ are known in terms of $m$, $T$ and $u\cdot p$ and can be 
expressed as
\begin{align}
B_2 =\,& \frac{1}{2J^{(0)}_{42}}, \label{B2}\\
\frac{D_0}{3B_0} =\,& 4\frac{J^{(0)}_{31}J^{(0)}_{20} - J^{(0)}_{41}J^{(0)}_{10}}
{J^{(0)}_{30}J^{(0)}_{10} - J^{(0)}_{20}J^{(0)}_{20}} \equiv C_2,\label{D0}\\
\frac{E_0}{3B_0} =\;& m^2 - 4\frac{J^{(0)}_{31}J^{(0)}_{30} - J^{(0)}_{41}J^{(0)}_{20}}
{J^{(0)}_{30}J^{(0)}_{10} - J^{(0)}_{20}J^{(0)}_{20}} \equiv C_1,\label{E0}\\
B_0 =\,& -\frac{1}{3C_1J^{(0)}_{21} + 3C_2J^{(0)}_{31} - 3J^{(0)}_{41} + 5J^{(0)}_{42}},\label{B0}
\end{align}
where the thermodynamic functions $J^{(r)}_{nq}$ are defined as
\begin{equation}\label{ICJ}
J_{nq}^{(r)} \equiv \frac{1}{(2q+1)!!}\!\int\! dP\, 
(u\cdot p)^{n-2q-r}\,(\Delta_{\mu\nu}p^\mu p^\nu)^q f_0\tilde f_0.
\end{equation}
In the above equation, the indices $n-r$ and $q$ represents the 
number of times $p^\mu$ and $\Delta_{\mu\nu}$ appear in the 
integration, respectively.

An analogous expression for $\delta f$ can also be obtained using an 
iterative Chapman-Enskog like solution of the relativistic Boltzmann 
equation in the relaxation-time approximation \cite 
{Romatschke:2011qp, Jaiswal:2014isa}. In absence of dissipation due 
to particle diffusion, the Chapman-Enskog method leads to \cite 
{Jaiswal:2014isa}
\begin{align}\label{DFCE}
\delta f_{\rm CE} = \frac{\beta }{u\cdot p}\bigg[&- \frac{1}{3\beta_\Pi}\left\{m^2-(1-3c_s^2)(u\cdot p)^2\right\}\Pi \nonumber\\
& + \frac{1}{2\beta_\pi}\;p^\mu p^\nu\pi_{\mu\nu}\bigg]f_0\tilde f_0,
\end{align}
where
\begin{align}
\beta_\Pi &= \frac{5}{3}\beta_\pi - (\epsilon+P)c_s^2, \label{betaPi}\\
\beta_\pi &= \beta\, J_{42}^{(1)}, \label{betapi}
\end{align}
and $c_s^2\equiv dP/d\epsilon$ is the speed of sound squared which 
can be expressed as
\begin{equation}\label{cs2}
c_s^2 = \frac{\epsilon+P}{\beta J_{30}^{(0)}}.
\end{equation}


\section{Viscous evolution equations}

The second-order evolution equations for $\Pi$ and $\pi^{\mu\nu}$ 
can be derived by considering the co-moving derivative of 
Eqs.~({\ref{BVP}) and ({\ref{SST}),
\begin{align}
\dot\Pi &= -\frac{1}{3}\Delta_{\alpha\beta} \!\int\! dP \, p^\alpha p^\beta\, \delta\dot f, \label{BVPE}\\
\dot\pi^{\langle\mu\nu\rangle} &= \Delta^{\mu\nu}_{\alpha\beta} \!\int\! dP \, p^\alpha p^\beta\, \delta\dot f. \label{SSTE}
\end{align}
Further, $\delta\dot f$ appearing in the above equations can be 
simplified by rewriting the relativistic Boltzmann equation,
\begin{equation}\label{RBE}
p^\mu\partial_\mu f = C[f],
\end{equation} 
in the form \cite{Denicol:2010xn}
\begin{equation}\label{RBER}
\delta\dot f = -\dot f_0 - \frac{1}{u\cdot p}\, p^\mu\nabla_\mu f + \frac{1}{u\cdot p}\, C[f].
\end{equation}
In the following, we consider the relaxation-time approximation for 
the collision term in the Boltzmann equation, 
\begin{equation}\label{RTAC}
C[f] = -\left(u\cdot p\right) \frac{\delta f}{\tau_{\rm eq}},
\end{equation}
where $\tau_{\rm eq}$ is the relaxation time. In general, in order 
for the collision term in the above equation to respect the 
conservation of particle four-current and energy-momentum tensor, 
$\tau_{\rm eq}$ must be independent of momenta and $u^\mu$ has to be 
defined in the Landau frame \cite{Anderson_Witting}.

Substituting $\delta\dot f$ from Eq.~(\ref{RBER}) into Eqs.~({\ref
{BVPE}) and ({\ref{SSTE}) along with the form of $\delta f$ given in 
Eqs.~({\ref{G14M}) and ({\ref{DFCE}), and after performing the 
integrations, we obtain
\begin{align}
\dot{\Pi} =& -\frac{\Pi}{\tau_{\rm eq}}
-\beta_{\Pi}\theta 
-\delta_{\Pi\Pi}\Pi\theta
+\lambda_{\Pi\pi}\pi^{\mu\nu}\sigma_{\mu \nu }, \label{BULK}\\
\dot{\pi}^{\langle\mu\nu\rangle} =& -\frac{\pi^{\mu\nu}}{\tau_{\rm eq}}
+2\beta_{\pi}\sigma^{\mu\nu}
+2\pi_{\gamma}^{\langle\mu}\omega^{\nu\rangle\gamma}
-\delta_{\pi\pi}\pi^{\mu\nu}\theta  \nonumber \\
&-\tau_{\pi\pi}\pi_{\gamma}^{\langle\mu}\sigma^{\nu\rangle\gamma}
+\lambda_{\pi\Pi}\Pi\sigma^{\mu\nu}, \label{SHEAR}
\end{align}
where $\omega ^{\mu \nu }\equiv \frac{1}{2}(\nabla^{\mu}u^{\nu} 
-\nabla^{\nu }u^{\mu })$ is the vorticity tensor. Note that the 
above form of the evolution equations for the bulk viscous pressure 
and the shear stress tensor is exactly same for both Grad's 
14-moment approximation ($\delta f_G$) and the Chapman-Enskog 
expansion ($\delta f_{CE}$). Moreover, the expressions for the first 
order transport coefficients, $\beta_{\Pi}$ and $\beta_{\pi}$, are 
also identical for these two cases and are given in Eqs.~(\ref
{betaPi}) and (\ref{betapi}). However, the second order transport 
coefficients appearing in the above equations are different for the 
Grad's 14-moment method and the Chapman-Enskog method.

The transport coefficients in the case of Grad's 14-moment 
approximation are calculated to be
\begin{align}
\delta^{\rm (G)}_{\Pi\Pi} &= 1 - c_s^2 - \frac{m^4}{9}\gamma^{(0)}_2, \label{Gcoeff1}\\
\lambda^{\rm (G)}_{\Pi\pi} &= \frac{1}{3} - c_s^2 + \frac{m^2}{3}\gamma^{(2)}_2, \label{Gcoeff2}\\
\delta^{\rm (G)}_{\pi\pi} &= \frac{4}{3} + \frac{m^2}{3}\gamma^{(2)}_2, \label{Gcoeff4}\\
\tau^{\rm (G)}_{\pi\pi} &= \frac{10}{7} + \frac{4m^2}{7}\gamma^{(2)}_2, \label{Gcoeff3}\\
\lambda^{\rm (G)}_{\pi\Pi} &= \frac{6}{5} - \frac{2m^4}{15}\gamma^{(0)}_2, \label{Gcoeff5}
\end{align}
where
\begin{align}
\gamma^{(0)}_2 &= (E_0+B_0m^2)J^{(0)}_{-20} + D_0J^{(0)}_{-10} - 4B_0J^{(0)}_{00}, \label{gamma0}\\
\gamma^{(2)}_2 &= \frac{J^{(0)}_{22}}{J^{(0)}_{42}}. \label{gamma2}
\end{align}
On the other hand, these transport coefficients in the case of 
Chapman-Enskog method are obtained as
\begin{align}
\delta^{\rm (CE)}_{\Pi\Pi} &= -\frac{5}{9}\,\chi - c_s^2, \label{CEcoeff1}\\
\lambda^{\rm (CE)}_{\Pi\pi} &= \frac{\beta}{3\beta_\pi}\!\left(7J_{63}^{(3)}+2J_{42}^{(1)}\right) - c_s^2, \label{CEcoeff2}\\
\delta^{\rm (CE)}_{\pi\pi} &= \frac{5}{3} + \frac{7\beta}{3\beta_\pi}\, J_{63}^{(3)},\label{CEcoeff4}\\
\tau^{\rm (CE)}_{\pi\pi} &= 2 + \frac{4\beta}{\beta_\pi}\,J_{63}^{(3)}, \label{CEcoeff3}\\
\lambda^{\rm (CE)}_{\pi\Pi} &= -\frac{2}{3}\chi, \label{CEcoeff5}
\end{align}
where
\begin{equation}
\chi = \frac{\beta}{\beta_\Pi}\!\left[(1-3c_s^2)\!\left(J_{42}^{(1)}+J_{31}^{(0)}\right) 
- m^2\!\left(J_{42}^{(3)}+J_{31}^{(2)}\right)\right]. 
\end{equation}

The integral functions $J_{nq}^{(r)}$ appearing in the 
expressions for the transport coefficients satisfy the relations
\begin{align}
J_{nq}^{(r)} & = \frac{1}{(2q+1)}\left[m^2 J_{n-2,q-1}^{(r)} - J_{n,q-1}^{(r)}\right], \label{prop1} \\
J_{nq}^{(0)}&= \frac{1}{\beta}\left[-I_{n-1,q-1}^{(0)} + (n-2q)I_{n-1,q}^{(0)} \right],\label{prop2}
\end{align}
where,
\begin{equation}\label{ICI}
I_{nq}^{(r)} \equiv \frac{1}{(2q+1)!!}\!\int\! dP\, 
(u\cdot p)^{n-2q-r}\,(\Delta_{\mu\nu}p^\mu p^\nu)^q f_0.
\end{equation}
Here we readily identify $I_{20}^{(0)}=\epsilon$ and 
$I_{21}^{(0)}=-P$. Using Eqs.~(\ref{prop1}) and (\ref{prop2}), we 
obtain the identities 
\begin{align}
J_{42}^{(0)} &= \frac{m^2}{5}J_{21}^{(0)} - \frac{1}{5}J_{41}^{(0)}, \label{J042}\\
J_{31}^{(0)} &= -\frac{1}{\beta}(\epsilon+P). \label{J031}
\end{align}
To compute all the transport coefficients, we also need to determine 
the integrals $J_{20}^{(0)}$, $J_{10}^{(0)}$, $J_{41}^{(0)}$, 
$J_{21}^{(0)}$, $J_{30}^{(0)}$, $J_{-20}^{(0)}$, $J_{-10}^{(0)}$, 
$J_{00}^{(0)}$, $J_{22}^{(0)}$, $J_{63}^{(3)}$, $J_{42}^{(1)}$, 
$J_{42}^{(3)}$, and $J_{31}^{(2)}$. In the following, we obtain 
expressions for these quantities in terms of modified Bessel 
functions of the second kind.


\section{Transport coefficients}

Let us first simplify our equilibrium distribution function. Using 
the result of summation of a infinite geometric progression,
\begin{equation}\label{GPS}
\frac{1}{1+x} = 1-x+x^2-x^3\cdots = \sum_{l=0}^{\infty}(-1)^lx^l, \quad \rm{for}~|x|<1,
\end{equation}
we obtain,
\begin{equation}\label{EDFGP}
f_0 = \frac{e^{-\beta (u\cdot p)}}{1+ae^{-\beta (u\cdot p)}} = \sum_{l=1}^{\infty}(-a)^{l-1}e^{-l\beta (u\cdot p)}.
\end{equation}
Similarly, using the result obtained after differentiating Eq. (\ref
{GPS}), we obtain
\begin{equation}\label{FTFGP}
f_0\tilde f_0 = \frac{e^{-\beta (u\cdot p)}}{(1+ae^{-\beta (u\cdot p)})^2} = \sum_{l=1}^{\infty}l(-a)^{l-1}e^{-l\beta (u\cdot p)}.
\end{equation}
Using the above results for $f_0$ and $f_0\tilde f_0$, the 
thermodynamic integrals $I_{nq}^{(r)}$ and $J_{nq}^{(r)}$ can be 
written as,
\begin{align}
I_{nq}^{(r)} =&\ \frac{gT^{n+2-r}z^{n+2-r}}{2\pi^2(2q+1)!!}\,(-1)^q\sum_{l=1}^{\infty}(-a)^{l-1}
\!\!\int_0^\infty\!\!\! d\theta\, \label{Inqr} \\
&\times(\cosh\theta)^{n-2q-r}(\sinh\theta)^{2q+2}\,\exp(-lz\cosh\theta), \nonumber\\
J_{nq}^{(r)} =&\ \frac{gT^{n+2-r}z^{n+2-r}}{2\pi^2(2q+1)!!}\,(-1)^q\sum_{l=1}^{\infty}l(-a)^{l-1}
\!\!\int_0^\infty\!\!\!\! d\theta\, \label{Jnqr} \\
&\times(\cosh\theta)^{n-2q-r}(\sinh\theta)^{2q+2}\,\exp(-lz\cosh\theta), \nonumber
\end{align}
where $z\equiv\beta m=m/T$.

The integral coefficients $I_{nq}^{(r)}$ and $J_{nq}^{(r)}$ can be 
expressed in terms of modified Bessel functions of the second kind. 
The integral representation of the relevant Bessel function is given 
by
\begin{equation}\label{Bessel}
K_n(z) = \!\int_0^\infty\! d\theta \cosh(n\theta)\, \exp(-z\cosh\theta).
\end{equation}
The pressure, energy density and $J_{30}^{(0)}$, required to 
calculate the velocity of sound, can be expressed in terms of Bessel 
functions using the above equation 
\begin{align}
P =&\, \frac{gT^4z^2}{2\pi^2}\!\sum_{l=1}^{\infty}\frac{1}{l^2}(-a)^{l-1}K_2(lz), \label{Prs_K}\\
\epsilon =&\, \frac{gT^4z^3}{2\pi^2}\!\sum_{l=1}^{\infty}\frac{1}{l}(-a)^{l-1}K_3(lz) - P, \label{eng_K}\\
J_{30}^{(0)} =&\, \frac{gT^5z^5}{32\pi^2}\!\sum_{l=1}^{\infty}l(-a)^{l-1}\Big[K_5(lz)+K_3(lz)-2K_1(lz)\Big]. \label{J300}
\end{align}
The integral functions $J_{nq}^{(r)}$ appearing in the transport 
coefficients obtained using the Grad's 14-moment method can be 
expressed as
\begin{align}\label{relint_G}
J_{21}^{(0)} =& -\frac{gT^4z^2}{2\pi^2}\!\sum_{l=1}^{\infty}\frac{1}{l}(-a)^{l-1}K_2, \nonumber\\
J_{20}^{(0)} =&\, \frac{gT^4z^3}{2\pi^2}\!\sum_{l=1}^{\infty}(-a)^{l-1}K_3 + J_{21}^{(0)}, \nonumber\\
J_{10}^{(0)} =&\, \frac{gT^3z^3}{8\pi^2}\!\sum_{l=1}^{\infty}l(-a)^{l-1}\Big[K_3-K_1\Big], \nonumber\\
J_{41}^{(0)} =& -\frac{gT^6z^6}{192\pi^2}\!\sum_{l=1}^{\infty}l(-a)^{l-1}\Big[K_6-2K_4-K_2+2K_0\Big], \nonumber\\
J_{-20}^{(0)} =&\, \frac{g}{2\pi^2}\!\sum_{l=1}^{\infty}l(-a)^{l-1}\Big[K_0-K_{i,2}\Big], \nonumber\\
J_{-10}^{(0)} =&\, \frac{gTz}{2\pi^2}\!\sum_{l=1}^{\infty}l(-a)^{l-1}\Big[K_1-K_{i,1}\Big], \nonumber\\
J_{22}^{(0)} =&\, \frac{gT^4z^4}{240\pi^2}\!\sum_{l=1}^{\infty}l(-a)^{l-1}\Big[K_4-8K_2+15K_0-8K_{i,2}\Big],  \nonumber\\
J_{00}^{(0)} =&\, \frac{gT^2z^2}{4\pi^2}\!\sum_{l=1}^{\infty}l(-a)^{l-1}\Big[K_2-K_0\Big],
\end{align}
where the $(lz)$-dependence of $K_n$ and $K_{i,n}$ is implicitly 
understood. The integral functions appearing in the transport 
coefficients obtained using the Chapman-Enskog method are
\begin{align}\label{relint_CE}
J_{63}^{(3)} =& -\frac{gT^5z^5}{3360\pi^2}\!\sum_{l=1}^{\infty}l(-a)^{l-1}\Big[K_5-11K_3+58K_1 \nonumber\\
&-64K_{i,1}+16K_{i,3}\Big], \nonumber\\
J_{42}^{(1)} =&\, \frac{gT^5z^5}{480\pi^2}\!\sum_{l=1}^{\infty}l(-a)^{l-1}\Big[K_5\!-7K_3\!+22K_1\!-16K_{i,1}\Big], \nonumber\\
J_{42}^{(3)} =&\, \frac{gT^3z^3}{120\pi^2}\!\sum_{l=1}^{\infty}l(-a)^{l-1}\Big[K_3\!-9K_1\!+12K_{i,1}\!-4K_{i,3}\Big], \nonumber\\
J_{31}^{(2)} =& -\frac{gT^3z^3}{24\pi^2}\!\sum_{l=1}^{\infty}l(-a)^{l-1}\Big[K_3-5K_1+4K_{i,1}\Big].
\end{align}
Here the function $K_{i,n}(lz)$ is defined by the integral
\begin{equation}\label{kin}
K_{i,n}(lz) = \!\int_0^\infty\! \frac{d\theta}{(\cosh\theta)^n}\,\exp(-lz\cosh\theta).
\end{equation}
Note that the subscript $i$ in the above function is not an index 
and just serves to distinguish it from the Bessel functions. 
$K_{i,n}$ satisfy the following recurrence relation:
\begin{equation}\label{kinkn}
\frac{d}{dz}K_{i,n}(lz) = -lK_{i,n-1}(lz),
\end{equation}
which can be expressed in the integral form:
\begin{equation}\label{kinkn}
K_{i,n}(lz) = K_{i,n}(0) - l\!\int_0^z\! K_{i,n-1}(lz') dz'.
\end{equation}
We observe that by matching $K_{i,0}(lz)=K_0(lz)$, the above 
recursion relation can be used to evaluate $K_{i,n}(lz)$ up to any 
$n$.

Armed with the above expressions for $J_{nq}^{(r)}$ in terms of 
series summation of the Bessel function, it is instructive to 
calculate the ratio between the coefficient of bulk viscosity and 
shear viscosity. In the relaxation-time approximation, this ratio is 
given by $\zeta/\eta = \beta_\Pi/\beta_\pi$ \cite{Denicol:2014vaa, 
Jaiswal:2014isa}. In the small-$z$ approximation, using the series 
expansion of the Bessel functions in powers of $z$, we obtain
\begin{equation}\label{zetabyeta}
\frac{\zeta}{\eta} = \Gamma\left(\frac{1}{3}-c_s^2\right)^2 + {\cal O}(z^5),
\end{equation}
where $\Gamma=75$ for Boltzmann distribution and $\Gamma=48$ for 
Fermi-Dirac distribution. However, in the case of Bose-Einstein 
distribution (remember $a=-1$), we get $\Gamma=-15+36/(1+a)$, and 
therefore it diverges as $\sim1/(1+a)$ up to the leading-order. It 
is interesting to note that the above expression is similar to the 
well known relation, $\zeta/\eta=15(1/3-c_s^2)^2$, derived by 
Weinberg \cite {Weinberg}. The difference in the proportionality 
constant may be attributed to the fact that here we have considered 
a single component system whereas, motivated by applications to 
cosmology, Weinberg considered a system composed of mixture of 
radiation and matter.

\begin{figure}[t]
\begin{center}
\includegraphics[width=\linewidth]{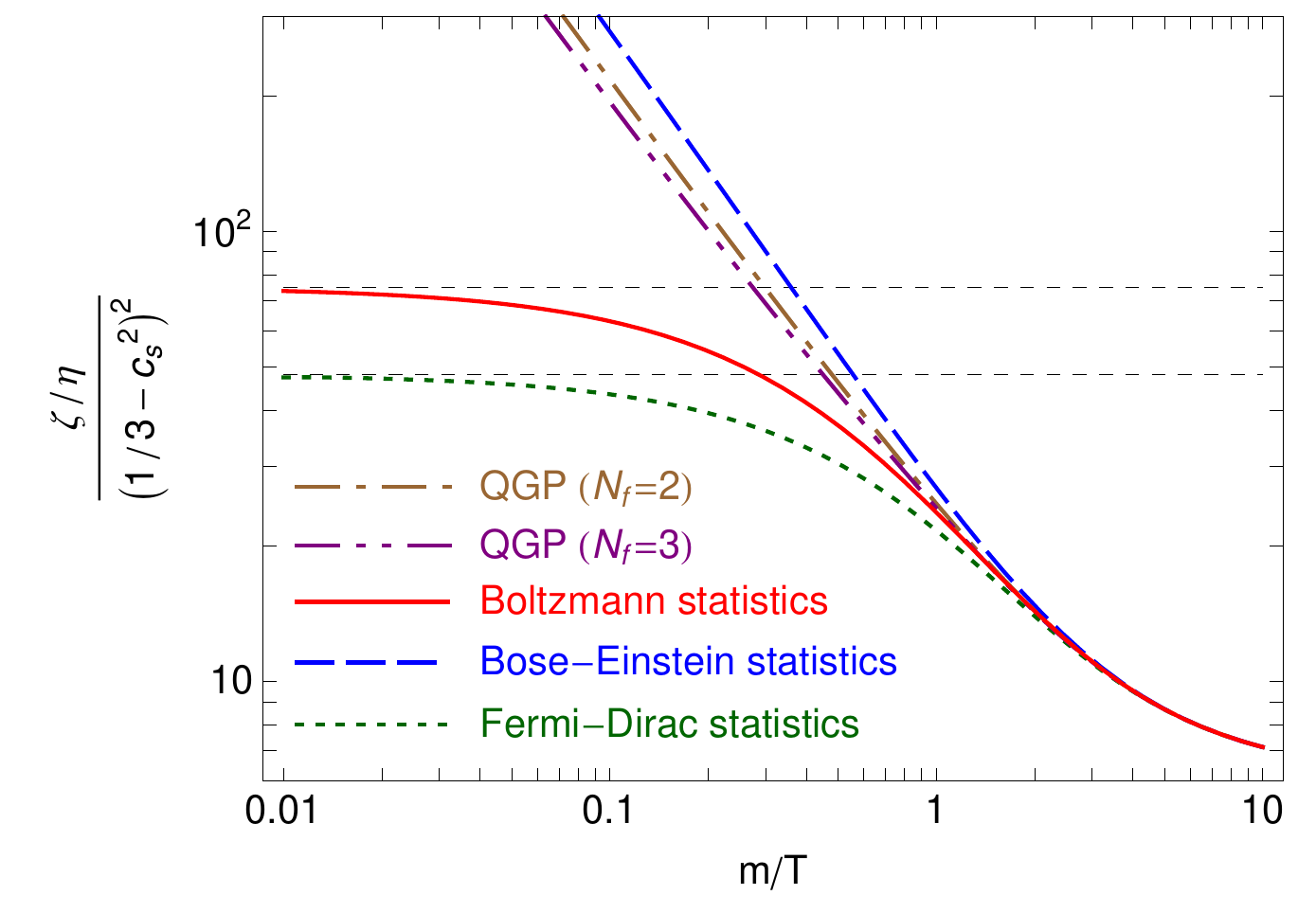}
\end{center}
\vspace{-0.8cm}
\caption{(Color online) $m/T$ dependence of the ratio $\zeta/\eta$ 
	scaled by $(1/3-c_s^2)^2$ for Boltzmann (red solid line), 
	Bose-Einstein (blue dashed line) and Fermi-Dirac (green dotted 
	line) statistics. We also show results for two-flavor (brown 
	dashed-dotted line) and three-flavor QGP (purple 
	dashed-dotted-dotted line). Thin horizontal black dashed lines 
	corresponds to constant values $75$ and $48$.} 
\label{ZetabyEta}
\end{figure}

In order to calculate $\zeta/\eta$ for the quark-gluon plasma (QGP), 
we need to provide appropriate degeneracy factors for the integral 
coefficients $I_{nq}^{(r)}$ and $J_{nq}^{(r)}$. For QGP, these 
integral coefficients will transform as
\begin{align}\label{IJnqr}
I_{nq}^{(r)} &\to I_{nq}^{(r)}|_{g=g_q,a=1} + I_{nq}^{(r)}|_{g=g_g,a=-1}, \nonumber\\
J_{nq}^{(r)} &\to J_{nq}^{(r)}|_{g=g_q,a=1} + J_{nq}^{(r)}|_{g=g_g,a=-1},
\end{align}
where $g_q$ and $g_g$ are the degeneracy factors corresponding to 
quarks and gluons respectively. These factors are given as
\begin{align}\label{degen}
g_q &= N_s \times N_{q\bar q} \times N_C \times N_f = 12N_f, \nonumber\\
g_g &= N_s \times \big(N_C^2 - 1\big) = 16,
\end{align}
where $N_f$ is the number of flavours, $N_C=3$ is the number of 
colours, $N_s=2$ corresponds to spin degrees and $N_{q\bar q}=2$ 
denotes quark and anti-quark.

\begin{figure}[t]
\begin{center}
\includegraphics[width=\linewidth]{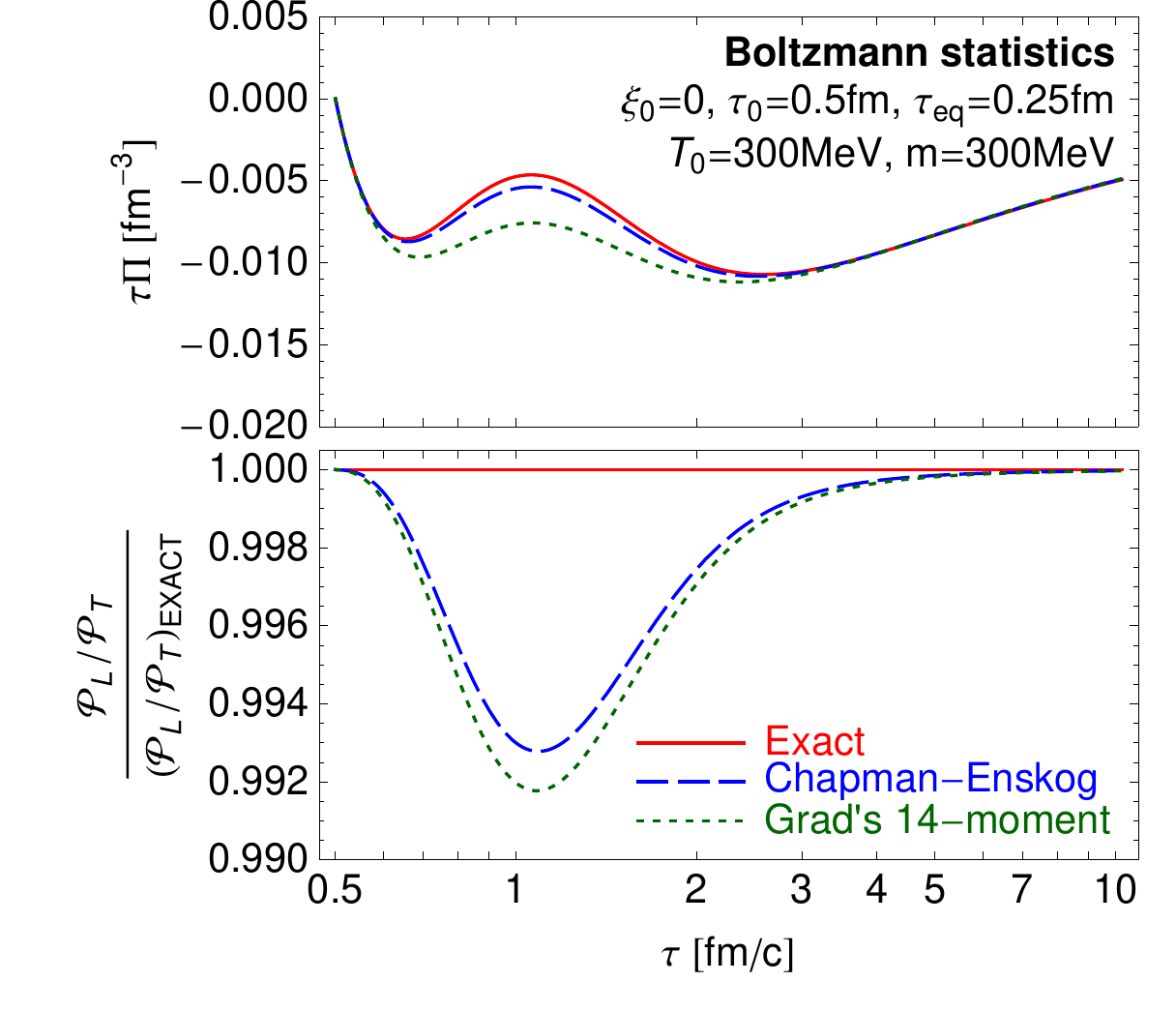}
\end{center}
\vspace{-0.8cm}
\caption{(Color online) Time evolution of the the bulk viscous 
	pressure times $\tau$ (top) and the pressure anisotropy ${\cal 
	P}_L/{\cal P}_T$ scaled by that obtained using exact solution of 
	the RTA Boltzmann equation (bottom) for Boltzmann statistics. 
	The three curves in both panels correspond to three different 
	calculations: the exact solution of the RTA Boltzmann equation 
	\cite{Florkowski:2014sda} (red solid line), second-order viscous 
	hydrodynamics obtained using the Chapman-Enskog method (blue 
	dashed line) and using the Grad's 14-moment approximation (green 
	dotted line). For both panels we use $T_0=300$ MeV at 
	$\tau_0=0.5$ fm/c, $m=300$ MeV, and $\tau_{\rm 
	eq}=\tau_\pi=\tau_\Pi=0.25$ fm/c. The initial spheroidal 
	anisotropy in the distribution function, $\xi_0=0$, corresponds 
	to isotropic initial pressures with $\Pi_0=\pi_0=0$.} 
\label{fig_0_B}
\end{figure}

In Fig.~\ref{ZetabyEta}, we show $m/T$ dependence of the ratio 
$\zeta/\eta$ scaled by $(1/3-c_s^2)^2$ for various cases. We observe 
that in accordance with the predictions from small-$z$ expansion, 
Eq.~(\ref{zetabyeta}), the curve for Boltzmann statistics (red solid 
line) and Fermi-Dirac statistics (green dotted line) saturates at 
$75$ and $48$ (thin horizontal black dashed lines), respectively, at 
small values of $z$. We also see that at very small-$z$, 
Bose-Einstein statistics (blue dashed line) results in very large 
values of the scaled viscosity ratio $(\zeta/\eta)/(1/3-c_s^2)^2$ 
indicating divergence. We see that this ratio for the QGP is 
dominated by the Bose-Einstein statistics for two-flavor (brown 
dashed-dotted line) as well as three-flavor QGP (purple 
dashed-dotted-dotted line). This however does not imply that the 
bulk viscosity of QGP is very large because a realistic calculation 
with lattice QCD equation of state suggest that $z>0.6$ even for 
temperatures as high as central RHIC and LHC energies \cite 
{Romatschke:2011qp, Andersen:2011sf, Strickland:2014zka}. Moreover, 
for large-$z$ we see that all three statistics lead to similar 
results for the scaled viscosity ratio which suggests that there is 
a relatively small effect coming from composition of the fluid.


\section{Boost-invariant 0+1d case}

In this section, we consider evolution in the case of transversely 
homogeneous and purely-longitudinal boost-invariant expansion \cite 
{Bjorken:1982qr}. For such an expansion all scalar functions of 
space and time depend only on the longitudinal proper time 
$\tau=\sqrt{t^2-z^2}$. Working in the Milne coordinate system, 
$(\tau,x,y,\eta)$, the hydrodynamic four-velocity is given by 
$u^\mu=(1,0,0,0)$. The energy-momentum conservation equation 
together with Eqs.~(\ref {BULK}) and (\ref{SHEAR}) reduce to
\begin{align}
\dot\epsilon &= -\frac{1}{\tau}\left(\epsilon + P + \Pi -\pi\right) \, ,  \label{epsBj}\\
\dot\Pi + \frac{\Pi}{\tau_\Pi} &= -\frac{\beta_\Pi}{\tau} - \delta_{\Pi\Pi}\frac{\Pi}{\tau}
+\lambda_{\Pi\pi}\frac{\pi}{\tau} \, ,  \label{bulkBj}\\
\dot\pi + \frac{\pi}{\tau_\pi} &= \frac{4}{3}\frac{\beta_\pi}{\tau} - \left( \frac{1}{3}\tau_{\pi\pi}
+\delta_{\pi\pi}\right)\frac{\pi}{\tau} + \frac{2}{3}\lambda_{\pi\Pi}\frac{\Pi}{\tau} \, , \label{shearBj}
\end{align}
where $\pi\equiv-\tau^2\pi^{\eta\eta}$. Note that in this case the 
term involving the vorticity tensor in Eq.~(\ref{SHEAR}) vanishes 
and hence has no effect on the dynamics of the fluid. Also note that 
the first terms on the right-hand side of Eqs.~(\ref{bulkBj}) and 
(\ref {shearBj}) corresponds to the first-order terms 
$\beta_\Pi\theta$ and $2\beta_\pi \sigma^{\mu\nu}$, respectively, 
whereas the rest are of second-order.

\begin{figure}[t]
\begin{center}
\includegraphics[width=\linewidth]{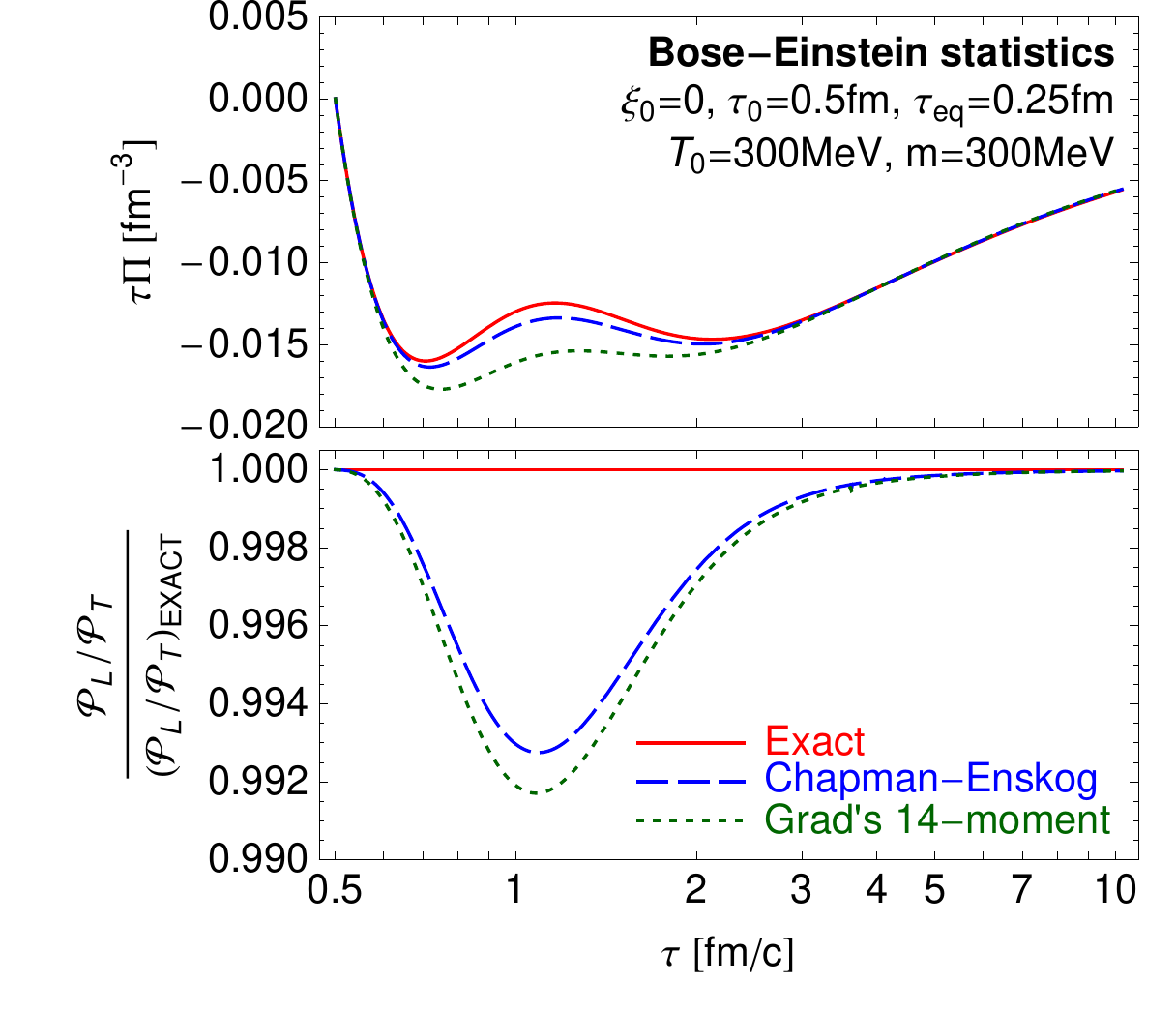}
\end{center}
\vspace{-0.8cm}
\caption{(Color online) Same as Fig. \ref{fig_0_B} except here 
	we consider Bose-Einstein statistics.} 
\label{fig_0_BE}
\end{figure}

\begin{figure}[t]
\begin{center}
\includegraphics[width=\linewidth]{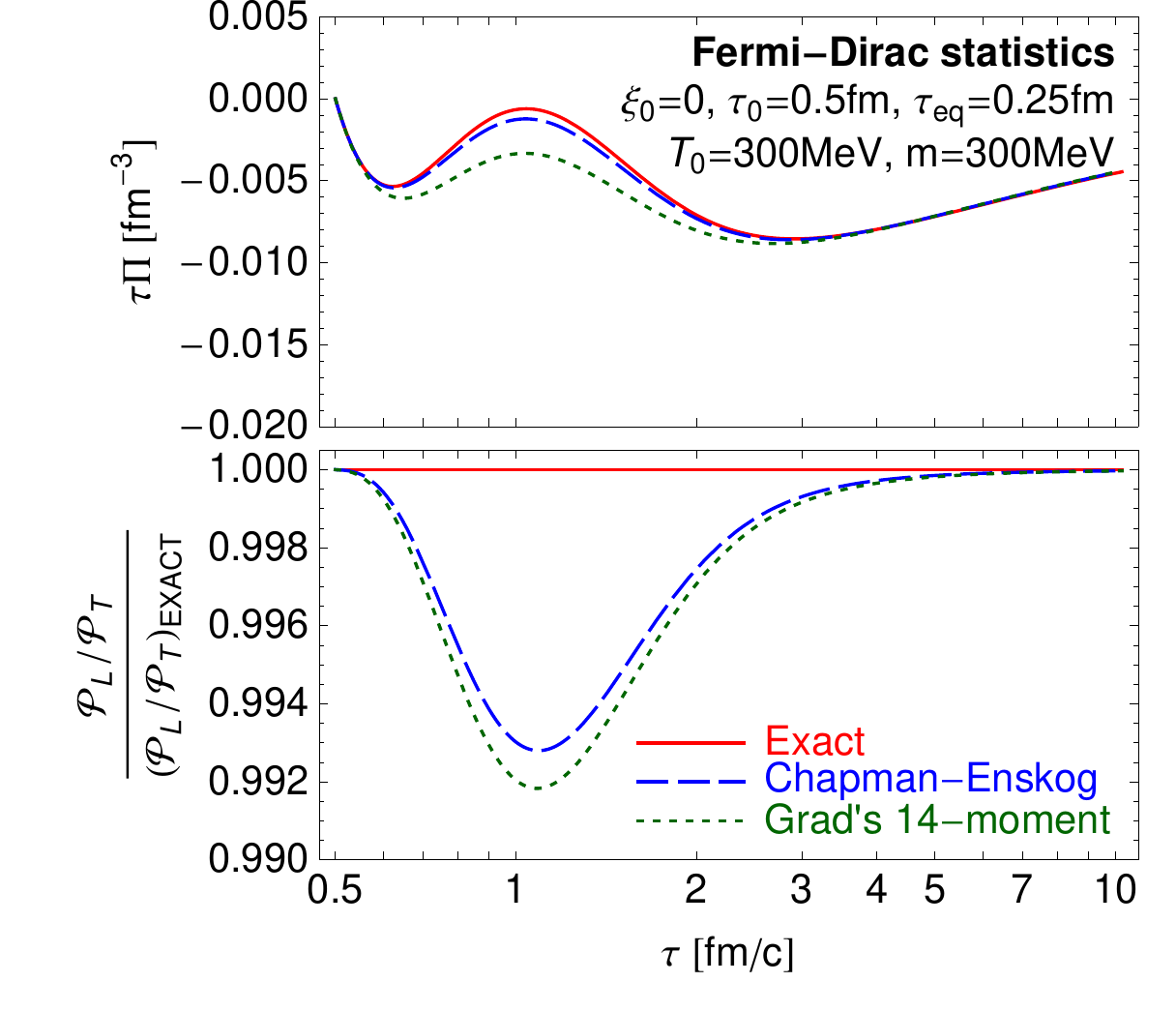}
\end{center}
\vspace{-0.8cm}
\caption{(Color online) Same as Fig. \ref{fig_0_B} except here 
	we consider Fermi-Dirac statistics.} 
\label{fig_0_FD}
\end{figure}

\begin{figure}[t]
\begin{center}
\includegraphics[width=\linewidth]{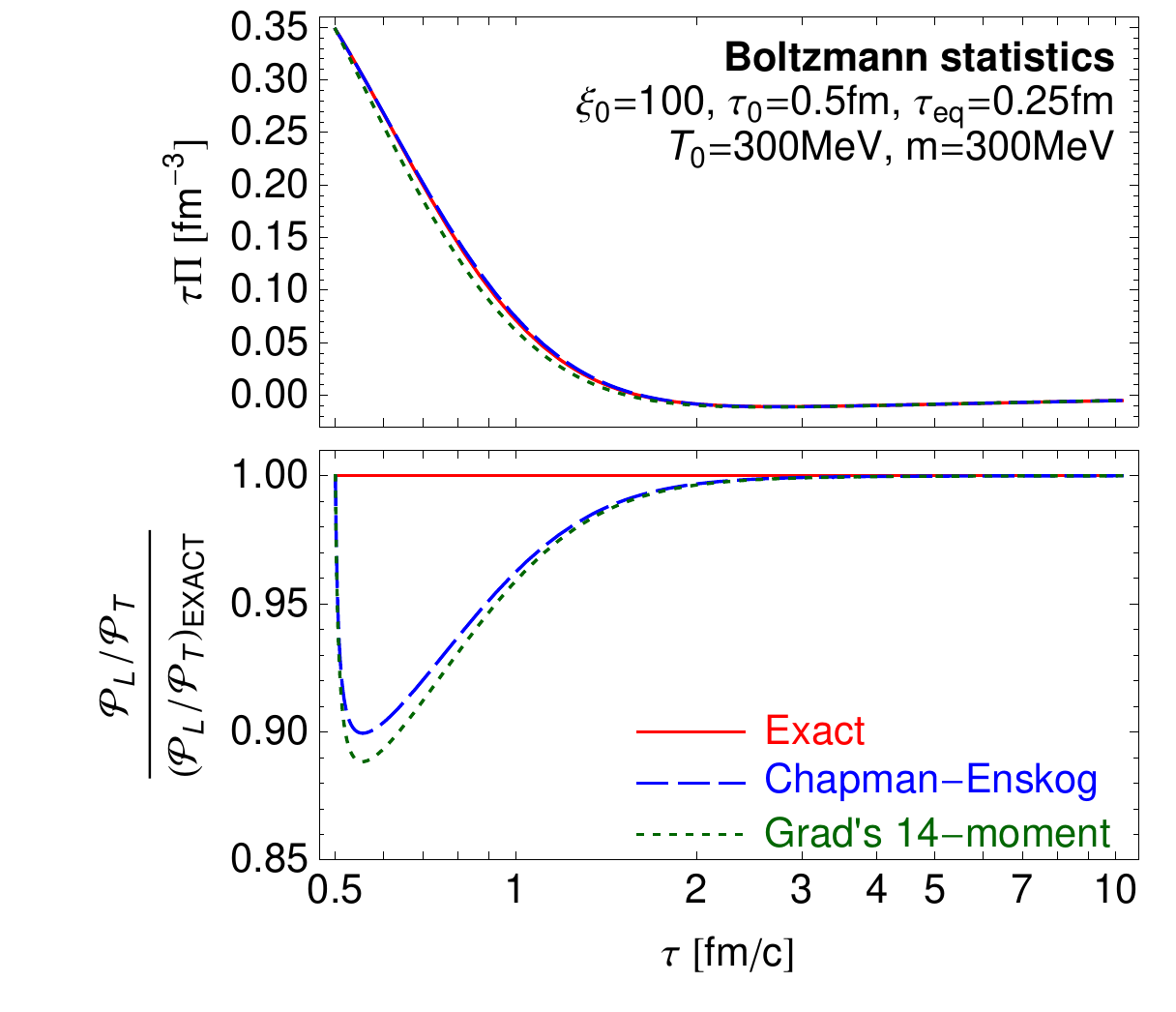}
\end{center}
\vspace{-0.8cm}
\caption{(Color online) Same as Fig. \ref{fig_0_B} except here 
	we take $\xi_0=100$ corresponding to $\pi_0=2.86$ GeV/fm$^3$ and 
	$\Pi_0=0.138$ GeV/fm$^3$.} 
\label{fig_100_B}
\end{figure}

\begin{figure}[t]
\begin{center}
\includegraphics[width=\linewidth]{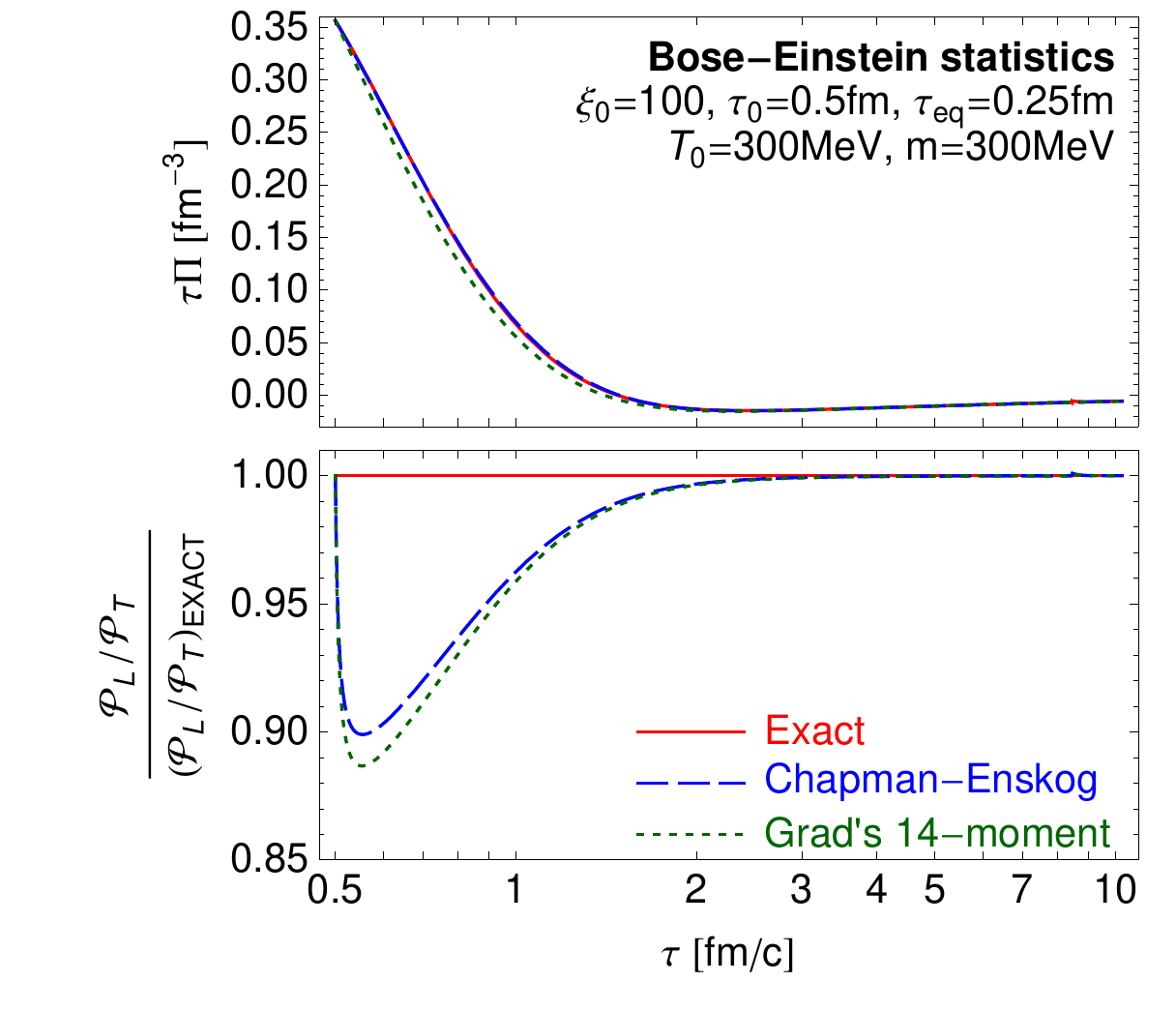}
\end{center}
\vspace{-0.8cm}
\caption{(Color online) Same as Fig. \ref{fig_100_B} except here 
	we consider Bose-Einstein statistics.} 
\label{fig_100_BE}
\end{figure}

\begin{figure}[t]
\begin{center}
\includegraphics[width=\linewidth]{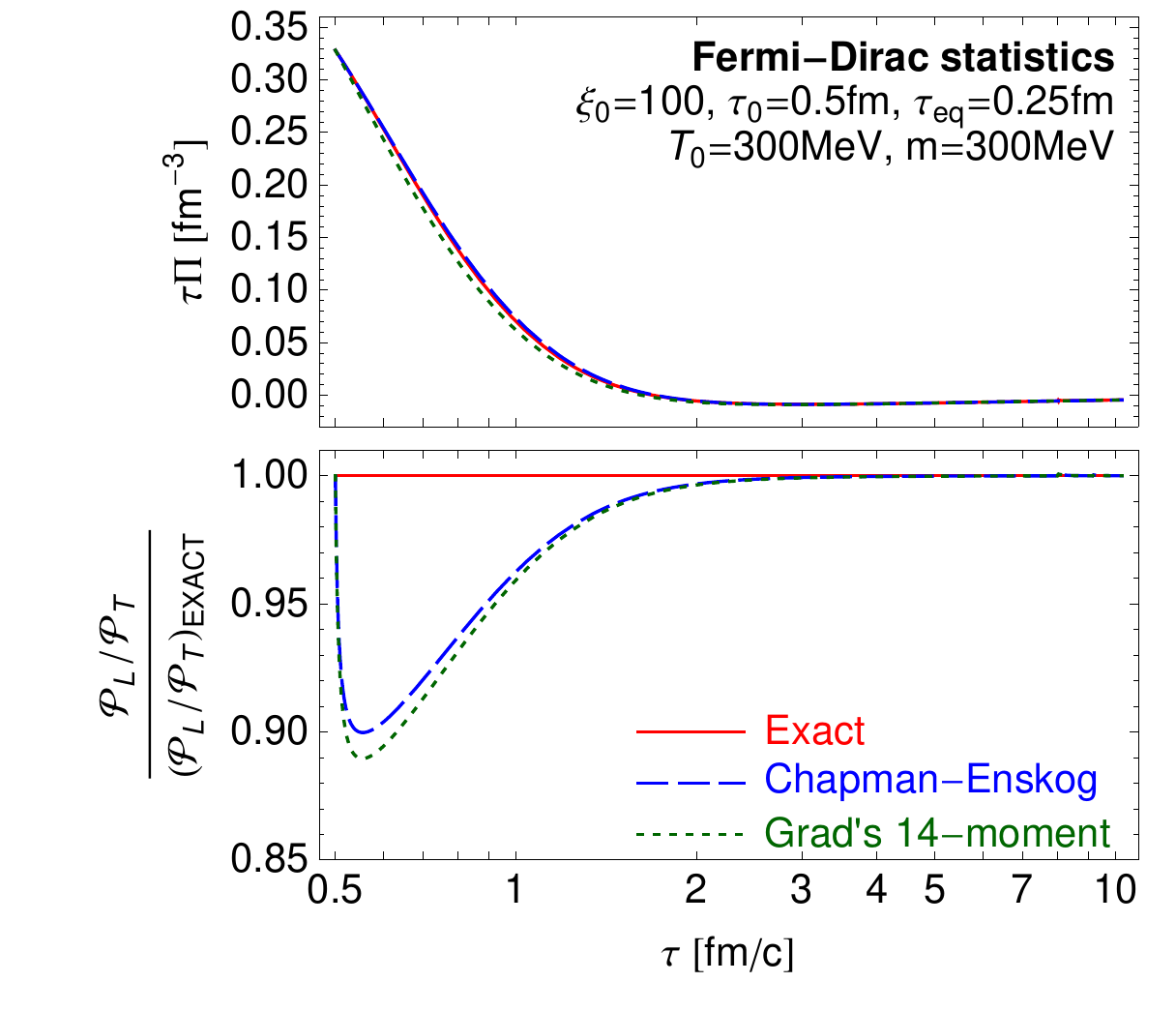}
\end{center}
\vspace{-0.8cm}
\caption{(Color online) Same as Fig. \ref{fig_100_B} except here 
	we consider Fermi-Dirac statistics.} 
\label{fig_100_FD}
\end{figure}

We simultaneously solve Eqs.~(\ref{epsBj})-(\ref{shearBj}) assuming 
an initial temperature of $T_0=300$ MeV at the initial proper time 
$\tau_0=0.5$ fm/c, with relaxation times $\tau_{\rm eq}=\tau_\Pi= 
\tau_\pi=0.25$ fm/c corresponding to initial $\eta/s=1/4\pi$. We 
solve these equations for two different initial pressure 
configurations: $\xi_0=0$ corresponding to an isotropic pressure 
configuration $\pi_0=\Pi_0=0$ and $\xi_0=100$ corresponding to a 
highly oblate anisotropic pressure configuration. The anisotropy 
parameter $\xi$ is related to the average longitudinal and 
transverse momentum in the local rest frame as: 
$\xi=\frac{1}{2}\langle p_T^2\rangle/\langle p_L^2\rangle-1$, where 
$\langle\cdots\rangle \equiv \int d^3p f_0(\sqrt{p_T^2 + (1+\xi) 
p_L^2},\Lambda)$ and $\Lambda$ is a temperature-like scale which can 
be identified with the temperature of the system in the isotropic 
equilibrium limit. For particle mass, we consider $m=300$ MeV which 
roughly corresponds to the constituent quark mass. We solve 
Eqs.~(\ref{epsBj})-(\ref {shearBj}) with transport coefficients 
obtained using the Grad's 14-moment method, 
Eqs.~(\ref {Gcoeff1})-(\ref{Gcoeff5}), as well as the Chapman-Enskog 
method, Eqs.~(\ref{CEcoeff1})-(\ref {CEcoeff5}).

In Figs.~\ref{fig_0_B} -- \ref{fig_100_FD} we show the time 
evolution of the bulk viscous pressure times proper-time (top) and 
the pressure anisotropy ${\cal P}_L/{\cal P}_T\equiv 
(P+\Pi-\pi)/(P+\Pi+\pi/2)$ scaled by that obtained using exact 
solution of the RTA Boltzmann equation (bottom) for three different 
calculations: the exact solution of the RTA Boltzmann equation \cite 
{Florkowski:2014sda} (red solid line), second-order viscous 
hydrodynamics obtained using the Chapman-Enskog method, Eqs.~(\ref 
{CEcoeff1})-(\ref {CEcoeff5}), (blue dashed line) and using the 
Grad's 14-moment approximation, Eqs.~(\ref{Gcoeff1})-(\ref 
{Gcoeff5}), (green dotted line). Figures~\ref {fig_0_B} and \ref 
{fig_100_B} are for Boltzmann statistics, Figs.~\ref{fig_0_BE} and 
\ref{fig_100_BE} are for Bose-Einstein statistics, and Figs.~\ref 
{fig_0_FD} and \ref{fig_100_FD} are for Fermi-Dirac statistics. 
Figs.~\ref {fig_0_B} -- \ref {fig_0_FD} correspond to an isotropic 
initial condition ($\xi_0=0$), while Figs.~\ref{fig_100_B} -- \ref 
{fig_100_FD} correspond to a highly oblate anisotropic initial 
condition ($\xi_0=100$).

From Figs.~\ref{fig_0_B} -- \ref{fig_100_FD}, we observe that, 
compared to Grad's 14-moment approximation, the transport 
coefficients obtained using the Chapman-Enskog method does a 
marginally better job in reproducing the ${\cal P}_L/{\cal P}_T$ 
obtained using the exact solution of the RTA Boltzmann equation. On 
the other hand, the result for $\tau\Pi$ obtained using the 
Chapman-Enskog method shows better agreement with the exact solution 
of the RTA Boltzmann equation than the Grad's 14-moment method.


\section{Conclusions and outlook}

In this paper we expressed the transport coefficients appearing in 
the second-order viscous hydrodynamical evolution of a massive gas 
using Bose-Einstein, Boltzmann and Fermi-Dirac statistics for the 
equilibrium distribution function and Grad's 14-moment approximation 
as well as the method of Chapman-Enskog expansion for the 
non-equilibrium part. The second-order viscous evolution equations 
are obtained by coarse graining the relativistic Boltzmann equation 
in the relaxation-time approximation. We also obtained the ratio of 
the coefficient of bulk viscosity to that of shear viscosity, in 
terms of the speed of sound, for classical and quantum statistics as 
well as for the QGP. We then considered the specific case of a 
transversally homogeneous and longitudinally boost-invariant system 
for which it is possible to exactly solve the RTA Boltzmann equation 
\cite{Florkowski:2014sda}. Using this solution as a benchmark, we 
compared the pressure anisotropy and bulk viscous pressure evolution 
obtained by employing both the Chapman-Enskog method as well as the 
Grad's 14-moment method. We demonstrated that the Chapman-Enskog 
method is in better agreement with the exact solution of the RTA 
Boltzmann equation compared to the Grad's 14-moment method. We found 
that, while both methods give similar results for the pressure 
anisotropy, the Chapman-Enskog method better reproduces the exact 
solution for the bulk viscous pressure evolution. 

At this juncture, we would like to clarify that we have used the 
exact solution of the Boltzmann equation, in the relaxation-time 
approximation, as a benchmark in order to compare different 
hydrodynamic formulations. The relaxation-time approximation is 
based on the assumption that the collisions tend to restore the 
phase-space distribution function to its equilibrium value 
exponentially. Although the microscopic interactions of the 
constituent particles are not captured in this approximation, it is 
reasonably accurate to describe a system which is close to local 
thermodynamic equilibrium \cite{Dusling:2009df}. Looking forward, it 
will be interesting to determine the impact of the quantum transport 
coefficients, obtained herein, in higher dimensional simulations. 
Moreover, it would also be instructive to see if the second-order 
results derived herein could be extended to obtain third order 
transport coefficients for quantum statistics \cite 
{Jaiswal:2013vta}. We leave these questions for a future work.


\begin{acknowledgments}

A.J. acknowledges useful discussions with Gabriel Denicol, Bengt 
Friman and Krzysztof Redlich. W.F. and E.M. were supported by Polish 
National Science Center Grant No.~DEC-2012/06/A/ST2/00390. A.J. was 
supported by the Frankfurt Institute for Advanced Studies (FIAS). 
R.R. was supported by Polish National Science Center Grant 
No.~DEC-2012/07/D/ST2/02125. M.S. was supported in part by U.S.~DOE 
Grant No.~DE-SC0004104.

\end{acknowledgments}


\end{document}